\documentclass[preprint,showpacs,preprintnumbers,amsmath,amssymb]{revtex4}

\usepackage{graphicx}
\usepackage{dcolumn}
\usepackage{bm}
\begin{document}

\title{Superstatistics in Econophysics}
\author{Yoshikazu Ohtaki}
\email{yoshi_oo@rf7.so-net.ne.jp}
\author{Hiroshi H. Hasegawa}
\email{hhh@mx.ibaraki.ac.jp}
\altaffiliation[Also at ]
{Center for Statistical Mechanics, University of Texas Austin, TX78712, USA.}
\affiliation{Department of Mathematical Sciences, Ibaraki University Mito 310-8512, Japan.}
\date{\today}

\begin{abstract}
We consider an ideal closed stock market, in which 100 traders have economic activities. 
The assets of the traders change through buying and selling stocks. 
We simulate the assets under conservation of both total currency and total number of stocks. 
If the traders are identical, then the assets are distributed as a stationary Gaussian. 
When variety among the traders makes winners and losers, 
the asset distribution displays power law scaling such as the Pareto law \cite{Pareto}. 
We discuss this power law scaling from the point of view of superstatistics \cite{Beck--Cohen}.
It is given as a superposition of scaled distributions for each hierarchical level. 
The various traders have the same growth rate distribution to keep the scaling.
\end{abstract}



\maketitle

More than 100 years ago, Pareto found the power law scaling in individual income distributions\cite{Pareto}.
How does the universal statistical law appear from complicated economic activities?
Many approaches have been proposed to answer this question
\cite{Stanley-Amaral-Buldyrev,Stanley-Amaral-Plerou,Okuyama-Takayasu,Bouchaud}.
Recently mean field dynamics with multiplicative noises explained the power law behavior
\cite{Takayasu--Sato--Takayasu,Aoyama-Nagahara,Aoyama,Mizuno-Takayasu-Takayasu}.
But individual characters were completely neglected in the mean field approach.
We reconsider the power law scaling from the point of view of superstatistics\cite{Beck--Cohen}. 
It is a result of the superposition of scaled typical statistical distributions
with respect to the characteristics.

We will introduce a model of the stock market to understand the behavior of asset distributions in this letter.
The model is an ideal closed stock market. There is no flow in or out of traders, cash or stocks.
The total number of traders, total cash and total number of stocks are kept constant.
The rules of trade are given by the extended threshold model\cite{Ohtaki--HHH}.
\par
 
In the comparison of finance and physics, we can find several common features \cite{Voit-book,Stanley_Book}.
In a stock market, a trader corresponds to a molecule in a gas system. 
Assets play the role of energy. The conservation of total assets means energy conservation. 
When traders are identical, we can expect the assets to become distributed as the Maxwell--Boltzmann.

A different behavior is expected when the characters of traders are widely divergent.
When there are superior and inferior traders in a stock market, 
some hierarchical levels such as poor, middle, and rich appear.
The asset distribution displays power-law scaling, 
which is different from the Maxwell--Boltzmann type distribution seen for identical traders.
As will be shown later, the power law scaling is realized as a result of superposition of 
scaled distributions for each hierarchical level.
\par

In this letter we define superstatistics as a superposition of scaled statistical distributions with 
respect to a characteristic parameter.
Our definition may be more general than the original one by Beck--Cohen\cite{Beck--Cohen}.
The idea of a superposition of scaled distributions has been developed in some fields\cite{magnet,Gruneis,Agu}.
Recently Beck and Cohen introduced superstatistics to explain the Tsallis statistics given as a superposition
of the Boltzmann distributions with respect to certain temperature distribution\cite{Beck--Cohen}. 
They also consider more general statistics for another temperature distributions.

From the aspect of superstatistics, the appearance of hierarchical structure is a key issue.
A trader stays at a fixed hierarchical level.
The possibility to be rich or poor beyond that hierarchical level is quite small.
This heterogeneous (non-ergodic) behavior is an important difference from the mean field approach.
We will discuss the relation between the two aspects in the end.

Now we consider an ideal closed stock market, in which $N$ traders buy and sell stocks. 
For simplicity, we choose only one brand of stock. 
The $i$th trader has currency $E_{i}(t)$ and stocks $V_{i}(t)$ at time $t$.
The stock market is completely isolated. 
There is no flow in or out of traders, currency or stocks. 
Therefore, the number of traders $N$, total currency $E_{\rm tot}=\sum_{i=1}^{N}E_{i}(t)$ and 
total number of stocks $V_{\rm tot}=\sum_{i=1}^{N}V_{i}(t)$ are kept constant.

From a statistical mechanical point of view, the assets of traders play the role of energy in the stock market. 
The ordinary assets of the $i$th trader is defined as $X_{i}\sim E_{i}(t)+V_{i}(t)p(t)$ where $p(t)$ is 
the price of stock at time $t$. 
In that definition, the total assets depend on time. 
In this letter we choose the following definition:
\begin{eqnarray}
X_{i}\equiv E_{i}(t)+V_{i}(t)p(0).
\end{eqnarray}
Although this definition is unusual, we give priority to strict conservation of total 
assets
\footnote{ Another way is to consider the following asset distribution rates instead of 
the assets themselves, $R_{i}(t)\equiv X_{i}(t)/X_{\rm tot}(t)$.
Then the total is always 1, since $\sum R_{i}=1$.}.

The asset of the $i$th trader changes through buying and selling stocks. 
In a trade at time $t$,
\begin{eqnarray}
\left. \begin{array}{l}
E_{i}(t) = E_{i}(t-1)  \pm \mu_{i}(t)p(t) \\
V_{i}(t) = V_{i}(t-t)  \mp \mu_{i}(t)
\end{array} \right.
\end{eqnarray}
where $\mu_{i}(t)$ is the volume of trade for the $i$th trader.
The buyer/seller pays/receives currency $\mu_{i}(t)p(t)$ to buy/sell $\mu_{i}(t)$ stocks.
Although the assets of the traders change, the total assets are kept constant.

Each trader demands/offers a bid/ask price and a buy/sell order volume at every time-step.
The bid and ask prices follow the rules introduced by Sato--Takayasu\cite{Sato-Takayasu}.
The Sato--Takayasu model is a deterministic threshold model \cite{Takayasu--Miura--Hirabayashi--Hamada}.
Although the model is very simple, its stock price has similar statistical properties as real stock prices
\footnote{ In the Sato--Takayasu model, the stock price is determined as 
$p(t)=(\max\{B_{i}(t)\}+\min\{S_{i}(t)\})/2$
when $\max\{B_{i}(t)\} \ge \min\{S_{i}(t)\}$. }.
\par

The bid/ask price of the $i$th trader is determined by the following rules:
\begin{eqnarray}
\left. \begin{array}{l}
B_{i}(t+1) = B_{i}(t) + a_{i}(t) + c_{i}[p(t)-p(t_{\rm prev}) ] \\
S_{i}(t+1) = B_{i}(t+1) + \Lambda_{i}
\end{array} \right.
\end{eqnarray}
where $\Lambda_{i}$, $|a_{i}|$ and $c_{i}$ are characteristic parameters of the $i$th trader.
The bid/ask price depends on the up/down movement of the stock price 
$c_{i}[p(t)-p(t_{\rm prev}) ]$ where $p(t_{\rm prev})$ is the stock price of the previous trade.
\par

The sign of $a_{i}(t)$ changes as $-a_{i}(t-1)$,  
when $a_{i}(t-1) > 0$ for the buyer or when $a_{i}(t-1) < 0$ for the seller.
Otherwise the sign is the same.
The bid/ask price changes $a_{i}$ in each step. 
After the $i$th trader becomes a buyer/seller, $a_{i}$ reverses its sign.
As a result, the $i$th trader alternately becomes a buyer and seller.
The period is approximately given as $\tau_{i} \sim 2\Lambda_{i}/|a_{i}(t)|$.
\par

The Sato--Takayasu model misses the volume of trade, which we need to estimate the asset distribution. 
We extended the threshold model to introduce a volume of trade\cite{Ohtaki--HHH}.
\par

We assume that a buy/sell order volume is proportional to the trader's assets.
\begin{eqnarray}
\left. \begin{array}{l}
\mu_{\rm b}(B_i)=\kappa E_{i}(t-1)/p(t-1)\\
\mu_{\rm s}(S_i)=\kappa V_{i}(t-1)
\end{array} \right.
\end{eqnarray}
where $\mu_{\rm b}(B_i)$/$\mu_{\rm s}(S_i)$ is a buy/sell order volume 
with the bid/ask price of the $i$th trader.
$\kappa$ is a constant. We choose $0<\kappa\le 1$.
This means that no trader can demand/offer a buy/sell order volume beyond 
their own assets.
\par

We determine the stock price and volume 
of trade using the following cumulative demand and offer functions\cite{Voit-book},
\begin{eqnarray}
\left. \begin{array}{l}
D(q)=\sum_{B_j\ge q}\mu_{\rm b}(B_i)\\
O(q)=\sum_{S_j\le q}\mu_{\rm s}(S_i)
\end{array} \right.
\end{eqnarray}
\par

The balance between the demand and offer determines the stock price and volume of trade.
(1) If $D(p*)=O(p*)$ for a unique $p*$, then the stock price is $p(t)=p*$.
If $D(p*)=O(p*)$ is unique, the volume of trade is $D(p*)=O(p*)$.
If $D(p*)=O(p*)$ is not unique,  the volume of trade is the maximum of $D(p*)=O(p*)$. 
(2)If $D(p*)=O(p*)$ for all $p*$ in $[p_{\rm min},p_{\rm max}]$, then the stock price is 
$p(t)=(p_{\rm min}+p_{\rm max})/2$. The volume of trade is $D(p*)=O(p*)$.
To satisfy the balance with respect to volume, we need to reduce some buy/sell order volumes.

\begin{figure}
\includegraphics[scale=0.7]{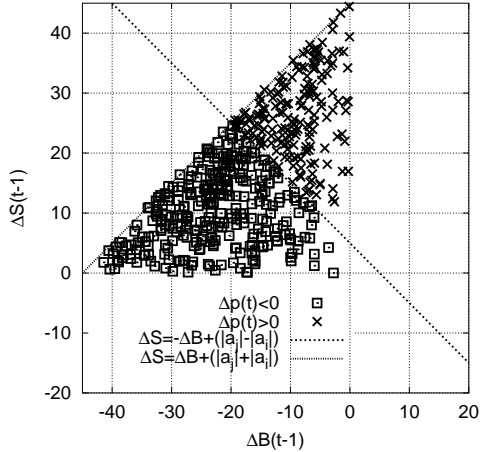}
\caption{\label{f.1} The large triangle with $\square$ and $\times$ represents the region in which Eqs.(7) are satisfied.
If $|a_{i}|<|a_{j}|$, the area of $\Delta p(t)>0~~ (\times)$ is less than $\Delta p(t)<0 ~~(\square)$.}
\end{figure}

The change of the bid/ask price of the $i$th trader, $|a_i|$, chracterizes the relative merit of 
the $i$th trader in the extended threshold model, since a trader with a smaller change can 
buy/sell stocks at a relatively lower/higher price.
We will discuss this property in
\footnote{ Suppose the $j$th trader sells some stocks and the $i$th trader buys them at time $t$.
The ask price of the $j$th trader is lower than the bid price of the $i$th trader: 
$B_i(t)=B_i(t-1)+|a_i|>S_j(t)=S_j(t-1)-|a_j|$.
Furthermore the stock price is lower than the ask price of the $j$th trader and 
lower than the bid price of $i$th trader at time $t-1$: $B_i(t-1)<p(t-1)<S_j(t-1)$.
From these relations, we obtain 
$\Delta B_i(t-1)<0$,
$\Delta S_j(t-1)>0$,
and $\Delta S_j(t-1)>\Delta B_i(t-1)+|a_i|+|a_j|$
where $\Delta B_i(t-1)\equiv B_i(t-1)-p(t-1)$ and $\Delta S_j(t-1)\equiv S_j(t-1)-p(t-1)$.
As is illustrated in Fig.\ref{f.1}, the region in which the above three conditions 
are satisfied is given as a triangle.
We assume that $\Delta B_i(t-1)$ and $\Delta S_j(t-1)$ are taken with equal probability within the triangle.
Since the stock price at time $t$ is given as $p(t)=(B_i(t)+S_j(t))/2$,
$\Delta p(t) \equiv p(t)-p(t-1)$
$=\frac{1}{2}(\Delta B_i(t-1)+\Delta S_j(t-1))+\frac{1}{2}(|a_i|-|a_j|)$.
For $\Delta p(t)>0$,
$\Delta S_j(t-1)>-\Delta B_i(t-1)-|a_i|+|a_j|$.
If $|a_i|<|a_j|$, the area of $\Delta p(t)>0$ is less than $\Delta p(t)<0$ in Fig.\ref{f.1}.
Therefore the $i$th trader can buy stocks relatively lower price. 
If $|a_i|>|a_j|$, the $j$th trader can sell stocks relatively higher price.}.
 
We simulated the time evolution of the stock market and obtained the asset distribution for long times.
We chose the following parameters and initial values:
$N=100$, $\Lambda_i = \Lambda = 1000$, $p_{0}=3000$, $V_{\rm tot}=10000$, 
$E_{\rm tot}=p_{0}V_{\rm tot}$, $\kappa=0.5$, $c_{i}=0$.
The initial bid price of the $i$th trader $B_{i}(0)$ is given as a random number within 
$(p_{0}-\Lambda,p_{0})$. 
The initial ask price of the $i$th trader is given as $S_{i}(0) = B_{i}(0) + \Lambda_{i}$.
Each trader has equal initial assets.
\par

We considered the following two typical cases with respect to the period of trade, $\tau_i=2\Lambda/|a_i|$
\begin{figure}
\includegraphics[scale=0.7]{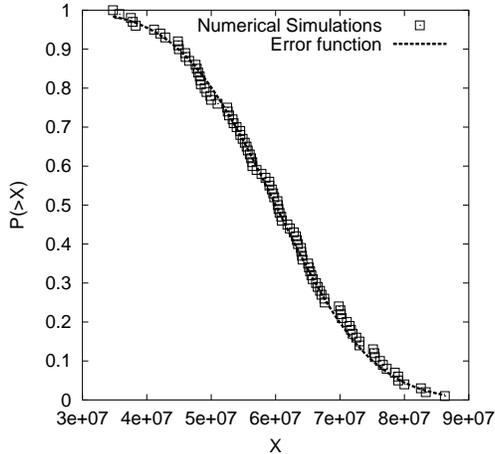}
\caption{\label{f.2_2} Cumulative distribution of assets for common $\tau$.
The dashed line is the error function.}
\end{figure}
\begin{figure}
\includegraphics[scale=0.7]{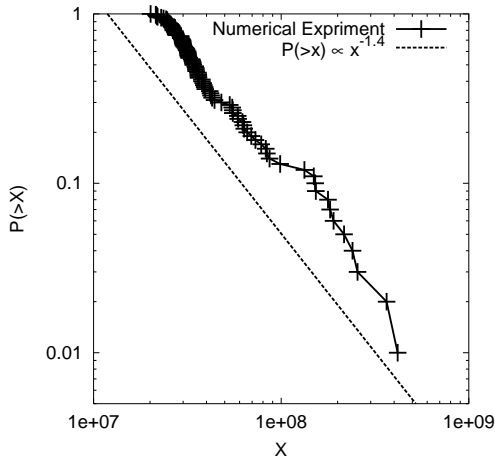}
\caption{\label{f.2} Cumulative distribution of assets for exponentially distributed $\tau$.
The dashed line is $X^{-1.4}$.}
\end{figure}
\par

\noindent{(i)Common $\tau_i~$}

In this case, the traders are completely identical. 
The period of trade is almost the same.
There are no strong/weak traders.
The mechanism of stability works in this financial market.
A trader who orders a greater/less volume has a tendency to lose/gain the asset.
Each trader gains and loses assets under the mechanism of stability. 
We numerically confirmed the assets to be distributed as a stationary Gaussian illustrated in Fig.\ref{f.2_2}. 
\par

\noindent{(ii)Exponentially distributed $\tau_i~$}

We chose $\tau_i$ as an exponentially distributed random number.
In this case, winners and losers appear. 
Although the distribution function cannot be completely steady,
it displayed the power law scaling illustrated in Fig.\ref{f.2}\footnote{
Although we discussed the power law scaling, the asset distribution is not completely steady.
This comes from the instability that 
rich traders have a tendency to gain big benefits even under the conservation of total assets.
The picture of equilibrium with small fluctuation is not valid.
There always exist large fluctuations.
To stabilize the distribution the role of tax becomes important.
We take some cash from rich traders and distribute them to poor traders.
We consider that a good tax keeps the scaling of distribution, and a bad one breaks it.
}.
\par

The assets were distributed as a stationary Gaussian for identical traders. 
For diverse traders, the asset distribution displayed the power law scaling. 
These results of our numerical simulations suggest that some hierarchical levels such 
as poor, middle, and rich appear. 
And the power law scaling is given as a superposition of scaled distributions for each hierarchical level. 
\begin{figure}
\includegraphics[scale=0.7]{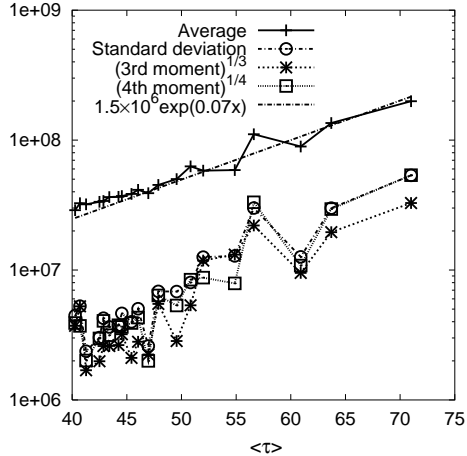}
\caption{\label{f.7} Average, standard deviation and higher order moments 
have the same exponent with respect to asset size.
The points for each moment are distinguished according to their asset size 
into twenty groups to clarify the size dependence.}
\end{figure}
\begin{figure}
\includegraphics[scale=0.7]{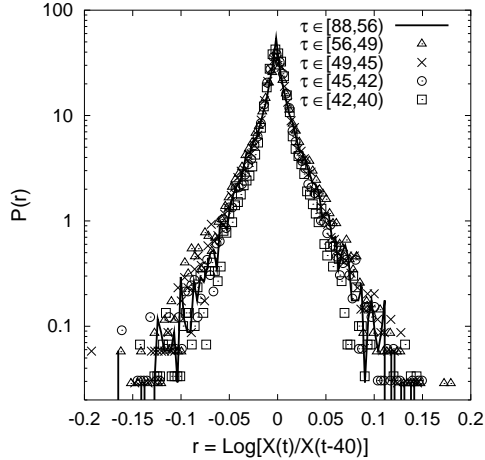}
\caption{\label{f.9} Probability density of growth rate $r \equiv \log(X(t)/X(t-40))$.
The points are distinguished according to their asset size into five groups.}
\end{figure}

We will explain how the power law appears from superstatistics for simple situations.
The relative merit of traders is characterized by the period of trade $\tau_i=2\Lambda/|a_i|$.
Traders who have longer period become winners.
For almost steady state, the time average of the assets of the $i$th trader is scaled as 
$m_i\sim\bar{m}\exp[\lambda\tau_i]$.
Furthermore the standard deviation of the time average of the assets is similarly scaled as 
$\sigma_i\sim\bar{\sigma}\exp[\lambda\tau_i]$.
This scaling may come from the fact that offer volume is always proportional to a trader's asset.
\par

If the time average of the assets of each trader is distributed as a Gaussian, 
the time averaged asset density $P(x)$ can be derived as
\begin{eqnarray}
\nonumber
P(x)\sim\sum_i f(\tau_i) \frac{1}{\sqrt{2\pi}\sigma_i}\exp[-\frac{(x-m_i)^2}{\sigma_i^2}]~~~~~~~~~ \\
\nonumber
\sim\int d\tau f(\tau) \frac{e^{-\lambda\tau}}{\sqrt{2\pi}\bar{\sigma}}\exp[-\frac{(x e^{-\lambda\tau}-\bar{m})^2}{\bar{\sigma}^2}]
\sim\frac{1}{x^{(1+\gamma/\lambda)}}
\end{eqnarray}
where $f(\tau_i)\sim e^{-\gamma\tau_i}$ is the probability density of the $i$th trader.
It is not necessary a Gaussian. Any scaled distributions are sufficient.

As was shown in Ref.\cite{Aoyama}, 
both scaled distribution and detailed balance are sufficient conditions for the Pareto law.
To confirm our approach we calculated the mean, the standard deviation and higher order moments.
These quantities showed scaling behavior as is shown in Fig.\ref{f.7}.
The Gibrat law (the size independence of the growth rate) is also illustrated in Fig.\ref{f.9}.
Notice that both the winners and losers have the same growth rate distribution to keep the Pareto law, even they have the different abilities.
These results are consistent with the theorem\cite{Aoyama,Fujiwara-Guilmi-Aoyama,Okuyama-Takayasu}.

In this letter, we have considered a model of the stock market in which 100 traders have economic activities. 
The assets of the traders change through buying and selling stocks. 
We have simulated the assets under the conservation of both total currency and total number of stocks. 
The assets become distributed as a stationary Gaussian for identical traders. 
When differences among traders make winners and losers, 
the asset distribution displays power law scaling such as the Pareto law. 
\par

We have interpreted this power law scaling from the point of view of superstatistics.
When the diversity of traders makes superiors and inferiors in a stock market, 
some hierarchical levels such as poor, middle, and rich appear. 
The asset distribution is given as a superposition of scaled distributions for each hierarchical level. 
\par
 
From the viewpoint of superstatistics, the appearance of hierarchical structure is a key issue.
A trader stays at a fixed level.
The possibility to be rich or poor beyond the corresponding level is quite small.
This heterogeneous (non-ergodic) behavior is an important difference from the approach of mean field dynamics.
\par

For an open market, the abilities of traders change in time.  Strong traders become weak and vice versa.
Only traders who can fit the present market become winners. 
Therefore the hierarchical structure becomes fuzzy for long time scales.
We expect that the mean field dynamics are recovered for such long time scales.
\par

For income distributions of companies, both the Pareto 
and Gibrat laws are valid in some well developed countries such as USA. and Japan
\cite{Aoyama,Fujiwara-Guilmi-Aoyama,Okuyama-Takayasu}.
As we have showed, the 
size independence of the growth rate does not deny the existence of 
hierarchical structures. The probability to be above or below the corresponding size is much smaller than what the mean field dynamics 
predicted. The hierarchical structures can be confirmed from empirical time 
series data. Growth rates can be related to the abilities of companies in quickly 
developing countries.

\begin{acknowledgments}
We are grateful to H. Yoon, M. Tanaka-Yamawaki, H. Takayasu and K. Nelson for useful discussions. 
We thank Y. Kondo and H. Matsumoto for their careful checking of program of the model. 
\end{acknowledgments}

\bibliography{reference}
\end{document}